\documentclass[aps,prc,twocolumn,superscriptaddress,showpacs,floatfix,nofootinbib]{revtex4-1}
\usepackage{color,graphicx}
\usepackage{dcolumn}
\usepackage{bm}
\usepackage{epsfig}
\usepackage{slashed}
\usepackage[bookmarksopen]{hyperref}
\def  \bea  {\begin{eqnarray}}
\def  \eea  {\end{eqnarray}}
\def  \nn   {\nonumber}

\def  \kperp   {k_{\perp}}

\def \bk {{\bf k}} 

\def \bq {{\bf q}}
\def \bp {{\bf p}}

\begin{document}

\title{Radiative heavy quark energy loss in an expanding viscous QCD plasma}

\author{Sreemoyee Sarkar}
\affiliation{Centre for Excellence in Basic Sciences, University of Mumbai,
Vidyanagari Campus, Mumbai 400098, India}
\author{Chandrodoy Chattopadhyay}
\affiliation{Department of Nuclear and Atomic Physics, Tata Institute of
Fundamental Research, Homi Bhabha Road, Mumbai 400005, India}
\author{Subrata Pal}
\affiliation{Department of Nuclear and Atomic Physics, Tata Institute of
Fundamental Research, Homi Bhabha Road, Mumbai 400005, India}

\begin{abstract}
We study viscous effects on heavy quark radiative energy loss in a dynamically screened medium
with boost-invariant longitudinal expansion. We calculate, to first order in opacity, the energy loss by 
incorporating viscous corrections in the single-particle phase-space distribution function within
relativistic dissipative hydrodynamics. We consider the Grad's 14-moment and the Chapman-Enskog-like
methods for the nonequilibrium distribution functions.
Our numerical results for the charm quark radiative energy loss show that, as compared to
static fluid, an expanding ideal (nonviscous) fluid causes a much smaller energy loss.
Viscosity in the evolution lead to somewhat enhanced energy loss which is insensitive to
the underlying viscous hydrodynamic models used. Further inclusion of viscous correction 
induces larger energy loss, the magnitude and pattern of this enhancement crucially
depend on the form of viscous corrections used.

\end{abstract}


\maketitle

\section{Introduction}

High-energy heavy-ion collision experiments at the Relativistic Heavy-Ion Collider (RHIC) 
\cite{Adams:2005dq,Adcox:2004mh} and the Large Hadron Collider (LHC) 
\cite{ALICE:2011ab,ATLAS:2012at,Chatrchyan:2013kba}
have firmly established the formation of strongly 
interacting matter composed of color deconfined system of quarks and gluons. 
Such conclusion was based from relativistic 
viscous hydrodynamic analysis of the large anisotropic flow that requires a remarkably 
small shear viscosity to entropy density ratio of $\eta/s = 0.08-0.20$ 
~\cite{Romatschke:2007mq,Schenke:2011bn,Qiu:2011hf,Bhalerao:2015iya}.
This not only suggests that the matter formed is close to local thermal equilibrium but also 
provides a window to the initial state of the fireball immediately after the collision. 
Hydrodynamic and transport model have been widely used to study the properties of
the hot and dense medium by exploring the collective flow of the soft (bulk) hadrons.

On the other hand, the suppression of high transverse momentum of light and heavy quarks
produced in hard processes provides an excellent tool that allows tomographic studies
of the QCD plasma 
\cite{Gyulassy:1991xb,Baier:1996sk,Lin:1997cn,Baier:2000mf,Gyulassy:1999zd,Djordjevic:2005db}. 
The suppression is caused by the attenuation (energy loss) of
the energetic partons via inelastic and elastic collisions during their propagation
in the medium. Heavy quarks, in particular, provide a promising probe as these are
formed in the early-stages via hard scatterings, and their production in the plasma 
at later stages is largely suppressed owing to their large mass 
\cite{Djordjevic:2005db}. These primordial
heavy quarks could thus explore various stages of the space-time evolution. 
While the energy loss is dominated by medium-induced gluon radiation at moderate and
large transverse momentum, collisional energy loss has sizable contribution 
especially at low transverse momentum $p_T$ \cite{Mustafa:2004dr}. 
However, the massive heavy flavor gives a reduced radiative energy 
loss at low $p_T$ (the ``dead cone effect") 
in contrast to light partons \cite{Dokshitzer:2001zm}. 
Consequently, RHIC measurements of the heavy flavor suppression 
data up to $p_T \approx 5$ GeV/c \cite{Adler:2005xv,elecQM05_STAR}
could be well reproduced in various distinct theoretical energy loss formalisms. 
Heavy ion collisions at LHC enable charm meson measurement at $p_T > 20$ GeV/c
and thus provide the ideal ground to study heavy flavor suppression
\cite{ALICE:2012ab,Alessandro:2014pfa}. In this
ultra-relativistic regime the radiative energy loss becomes dominant.
In contrast to pQCD predictions, a surprisingly large suppression pattern of high-$p_T$ 
D-mesons was observed at LHC, quantified by the nuclear modification
factor, $R_{AA} = (dN/dp_T)_{AA}/N_{\rm bin}(dN/dp_T)_{pp}$, defined as the ratio
of the yield in $AA$ and $pp$ collisions scaled by binary nucleon-nucleon collisions.
The crucial ingredient for a reliable suppression prediction relies on a precise
energy loss calculation taking due consideration of the expansion and viscosity 
of the QCD medium.

Early calculations of the medium-induced radiative energy loss were based on ``static
QCD medium" consisting of randomly distributed static scattering centers. In such a
static medium, the collisional energy loss exactly vanishes. Subsequently, the radiative
energy loss in a dynamically screened QCD medium was developed for an optically thick
plasma \cite{Djordjevic:2007at} and finite-size optically thin plasma 
\cite{Djordjevic:2008iz,Djordjevic:2009cr}, more relevant for
rapidly expanding medium formed in relativistic heavy ion collisions. 
Radiative energy loss in a plasma was also shown to receive corrections because 
of modified dielectric effect of the medium known as Ter-Mikayelian effect
\cite{Djordjevic:2003be}. Further, calculation of radiative energy loss suffers complication due to 
Landau-Pomeranchuck-Migdal (LPM) effect \cite{Migdal:1956tc,MVWZ:2004,Baier:1998kq} 
which introduces a controlled reduction of emitted gluon formation time.
As the increase of quark mass increases the phase-shifts and reduces the destructive nature of LPM,
the effect was found prominent for light quarks and gluons 
\cite{Gyulassy:1991xb,Baier:2000mf,Gyulassy:1999zd,Gyulassy:2003mc,TOMO}.

The existing calculations on energy loss have been performed purely for an ideal 
fluid using the equilibrium phase-space distribution function, and ignoring viscous effects.
Since the QGP formed in relativistic heavy-ion collisions behaves like a 
near-prefect fluid with a small $\eta/s$, it is imperative to account for the
viscous effects in computing the radiative energy loss. In fact, the importance of 
viscosity of the medium has already been realized in several quantities/observables 
relevant for RHIC and LHC, such as heavy quark damping rate \cite{Sarkar:2012fk}, 
anisotropic flow \cite{Romatschke:2007mq,Schenke:2011bn,Qiu:2011hf,Bhalerao:2015iya},
event-plane correlations \cite{Qiu:2012uy,Chattopadhyay:2017bjs}, 
dilepton spectra \cite{Dusling:2008xj,Vujanovic:2013jpa} etc. 
While few calculations have incorporated radiative energy loss only for
viscous medium evolution \cite{Baier:2000mf,Baier:2006pt}, a realistic 
and consistent calculation where viscosity is explicitly included in the computation 
of energy loss as well as in the expanding viscous medium is crucial.

In this paper, we present the first calculation of radiative energy loss with viscosity,
in first-order in opacity, of heavy (charm) quark in a dynamically screened viscous QCD medium 
that undergoes boost-invariant longitudinal expansion. We employ causal second-order viscous 
hydrodynamics for the underlying evolution of the medium based on the M\"uller-Israel-Stewart (MIS) 
framework \cite{Muller:1967zza,Israel:1979wp,Muronga:2003ta}
and the recently derived dissipative equations from Chapman-Enskog-like approach of iteratively solving 
the Boltzmann equation in the relaxation time approximation 
\cite{Jaiswal:2013npa,Bhalerao:2013pza,Chattopadhyay:2014lya}.
Viscous effects are incorporated in the single-particle distribution 
$f(x,p) = f_0(x,p) + \delta f_{\rm vis}(x,p)$, via the
nonequilibrium distribution function $\delta f_{\rm vis}$. 
The single-particle distribution would modify the scattering cross section of the energetic 
parton with the medium and thereby the radiative energy loss. 
For the nonequilibrium distribution, we use the commonly
used form based on Grad's 14-moment approximation 
\cite{Grad} and that obtained from Chapman-Enskog method.
We shall show that viscosity, in general, enhances the energy loss, 
the enhancement is significant in the Grad's method.  
However, the magnitude of the total energy loss with viscosity
included both in the dynamics and phase space distribution is smaller than 
that in the static limit. Further, we find nonlinearity in the time dependence 
of radiative energy loss for a viscous plasma, 
which mimics the energy loss behavior due to coherent gluon radiation. 

The paper is organized as follows. In Sec. II we introduce the dissipative hydrodynamic
formalisms used and then compute in these models, to first order in opacity, 
the radiative energy loss in a dynamical viscous QCD medium with boost-invariant 
longitudinal expansion. In Sec. III we compare the results for radiative energy loss 
in ideal and viscous fluids and with further inclusion of viscous corrections due
to Grad and Chapman-Enskog methods, by using 
initial conditions relevant to that produced in heavy-ion collisions at LHC. 
In Sec. IV we summarize and conclude. The technical details of the 
computation of energy loss are presented in Appendixes A-C.

\section{Radiative energy loss in an expanding viscous medium}      

In this section we compute the medium induced heavy flavor radiate energy loss in 
the boost-invariant longitudinal expansion of matter within second-order viscous hydrodynamics.
The hydrodynamic evolution is governed by the conservation of
energy-momentum tensor, $\partial_\mu T^{\mu\nu} =0$, where
\bea
T^{\mu\nu} = \epsilon u^\mu u^\nu - P\Delta^{\mu \nu} + \pi^{\mu\nu}.
\label{NTD:eq}
\eea
We shall work in the Landau-Lifshitz frame and disregard particle flow
due to very small values of net-baryon number formed at RHIC and LHC.
In the above equation,  $\epsilon$ and $P$ are respectively the energy density
and pressure in the fluid's local rest frame (LRF), and $\pi^{\mu\nu}$ is the shear
pressure tensor. $\Delta^{\mu\nu}=g^{\mu\nu}-u^\mu u^\nu$ is the
projection operator on the three-space orthogonal to the
hydrodynamic four-velocity $u^\mu$ defined by the Landau-matching
condition $T^{\mu\nu}u_\nu = \epsilon u^\mu$. 

For Bjorken longitudinal expansion, we work in the Milne coordinates ($\tau,x,y,\eta_s$)
where proper time $\tau = \sqrt{t^2-z^2}$, space-time rapidity $\eta_s = \ln[(t+z)/(t-z)]/2$,
and four-velocity $u^\mu = (1,0,0,0)$.
The conservation equation for the energy-momentum tensor gives the evolution of $\epsilon$: 
\bea
\frac{d\epsilon}{d\tau} = -\frac{1}{\tau} \left(\epsilon + P - \Phi \right),
\label{energy:eq}
\eea
where $\Phi \equiv -\tau^2\pi^{\eta_s\eta_s}$ is taken as the independent component
of the shear pressure tensor.
For the three independent variables, we need two more equations, namely, the viscous
evolution equation and the equation of state (EoS). In this work, we have used a conformal 
QGP fluid EoS with thermodynamic pressure $P = \epsilon/3$.
The simplest choice for the dissipative equation would be the relativistic 
Navier-Stokes theory, where the instantaneous constituent equation for the shear
pressure in the Bjorken case gives 
\bea
\Phi = \frac{4\eta}{3}\theta .
\label{NS:eq}
\eea
Here $\eta \geq 0$ is the shear viscosity coefficient and the local expansion rate
$\theta = 1/\tau$. However, this first-order theory suffers from acausality and instability.

The most commonly used second-order dissipative hydrodynamic equation, derived from positivity
of the divergence of entropy four-current, is based on the works of M\"uller-Israel-Stewart (MIS)
\cite{Muller:1967zza,Israel:1979wp,Muronga:2003ta}.
In the boost-invariant scaling expansion, the MIS dissipative equation,
\bea
\frac{d\Phi}{d\tau} + \frac{\Phi}{\tau_\pi} 
= \frac{4\eta}{3\tau_\pi} \theta - \lambda_\pi\theta\Phi ,
\label{MIS:eq}
\eea
restores causality by enforcing the shear pressure to relax to its first-order value via the
relaxation time $\tau_\pi = 2\eta \beta_2$, where $\beta_2$ is the
second-order transport coefficient. In the present study we consider 
$\tau_\pi = 2\eta \beta_2 = 5\eta/(sT)$ corresponding to that obtained in a 
weakly coupled QCD.  Further, the coefficient of the second-order term
(in the expansion of the velocity gradients) for EoS of an ultrarelativistic gas 
is $\lambda_\pi = 4/3$. 
In the derivation of Eq. (\ref{MIS:eq}) pertaining to a system that is out of equilibrium, 
the nonequilibrium effects have been quantified via the phase-space distribution,
$f(x,p) = f_0(x,p) + \delta f_{\rm vis}(x,p)$, where
the nonequilibrium part of the distribution function, $\delta f_{\rm vis}(x,p)$, is usually 
obtained by expanding $f(x,p)$ about the equilibrium distribution function 
$f_0(x,p) \approx [{\rm exp}(u \cdot p/T) -1]^{-1}$.
The Grad's 14-moment approximation \cite{Grad} is the common choice of viscous
correction in hydrodynamics, where the expansion in powers of momenta is truncated 
at quadratic order. For a system of massless
particles in the absence of bulk viscosity and charge diffusion current, the
Grad's method for viscous correction gives 
\bea
\delta f_{\rm vis} &=& f_0(1\pm f_0)\frac{\pi_{\mu\nu}p^{\mu}p^{\nu}}{2(\epsilon+P)T^2} , \nn\\
&\equiv& f_0(1\pm f_0)\frac{3\Phi}{4(\epsilon+P)T^2}\left(\frac{\vec \bp^2}{3} - p_z^2 \right) ,
\label{fvis_Grad:eq}
\eea
where the second line is the equivalent representation for boost-invariant longitudinal expansion 
(assumed to be along $z$-direction) of the fluid in the LRF \cite{Moore:2004tg}.
This equation has been exclusively used in deriving the second-order MIS dissipative equation.

Alternatively, dissipative evolution equations can be obtained from
Chapman-Enskog-like (CE) method by perturbative expansion of the Boltzmann
transport equation using Knudsen number as a small expansion parameter 
\cite{Jaiswal:2013npa,Bhalerao:2013pza,Chattopadhyay:2014lya}. By expanding the 
nonequilibrium distribution function $\delta f_{\rm vis}$ about the local equilibrium 
value, and iteratively solving the Boltzmann equation in the relaxation-time approximation,
the second-order dissipative equation for the shear tensor in the boost-invariant gives 
\bea
\frac{d\Phi}{d\tau} + \frac{\Phi}{\tau_\pi} 
= \frac{4\eta}{3\tau_\pi} \theta - \lambda_\pi\theta\Phi.
\label{CE:eq}
\eea
In the Chapman-Enskog-like approach, the relaxation time naturally comes out to be
$\tau_\pi = 2\eta\beta_2 = 5\eta/(s T)$ and $\lambda_\pi = 38/21$ \cite{Jaiswal:2013npa}.
The corresponding nonequilibrium distribution function has the form
\bea
\delta f_{\rm vis} &=& f_0(1 \pm f_0)\frac{5\pi_{\mu\nu} p^\mu p^\nu}{8PT(u\cdot p)} \nn\\
&\equiv & f_0(1 \pm f_0)\frac{15\Phi}{16PT(u\cdot p)} \left(\frac{\vec \bp^2}{3} - p_z^2 \right).
\label{fvis_CE:eq}
\eea
We shall present the calculational details of heavy quark radiative energy loss for the 
M\"uller-Israel-Stewart dissipative hydrodynamics with Grad's form of $\delta f_{\rm vis}(x,p)$. 
The results within the Chapman-Enskog approach can be obtained in a similar fashion, which will 
be presented at the end of this section.

We shall compute the energy loss in a dynamical QCD medium for a thick expanding plasma 
in the opacity expansion. In a static plasma the energy loss is calculated by expansion 
over the number of parton scatterers in the medium times the transport cross section, 
integrated over the path length $L$ traversed by the heavy quark. In an expanding 
medium, the total energy loss is obtained by summing the instantaneous energy loss
over the time spent by the quark in the plasma before reaching vacuum or the
survival time in the plasma.

In principle, boost invariance expansion induces anisotropy in the medium,
hence the energy lost by a quark depends on its direction of propagation 
relative to the fluid flow \cite{Romatschke:2004au}. 
In this paper, we consider the propagation of the
heavy quark to be along the fluid direction and relegate to future work the calculation of 
complicated directional dependence of energy loss.
As in Ref. \cite{Djordjevic:2007at,Djordjevic:2008iz}, 
we restrict ourselves to first order in opacity,
where an on-shell heavy quark of mass $M$ and spatial momentum $p \gg M$ produced in the
remote past traverses along fluid flow, i.e. $z$-direction. On scattering with a parton 
in the medium, it exchanges a virtual gluon of 
momentum $q = (q_0, \vec\bq) = (q_0, q_z, \bq)$ and radiates a gluon with 
momentum $k = (\omega, \vec\bk) = (\omega, k_z, \bk)$. The heavy quark then emerges along
the $z$-direction with a momentum $p' = (E', \vec\bp') = (E', p'_z, \bp')$. As the gluon
momentum is spacelike ($q_0 \leq |\vec\bq|$) and the radiated gluon momentum 
is timelike ($\omega \geq |\vec\bk|$), these contribute accordingly in the gluon
propagators $D^{\mu\nu}(q)$ and $D^{\mu\nu}(k)$, respectively. 
The validity of soft gluon, ($\omega \ll E$), and soft rescattering, 
($|{\bf q}| \sim |{\bf k}| \ll k_z$), approximations at high temperature $T$ at LHC
together with the energy-momentum conservation, $p = p' + k + q$, enables us to write 
\bea
k&=& \left( \omega \approx k_z\!+ \!\frac{\bk^2+m_g^2}{2 k_z},k_z, \bk \right),\nn\\
p'&=& \left( E'\approx p_z'\!+ \!\frac{\bp'^2 + M^2}{2 p'_z}, p_z', -(\bk+\bq) \right),\nn\\
p&=& \left( E\approx p_z' \!+ \!k_z\!+ \!q_z \!+ \!\frac{M^2}{2(p_z'+k_z+q_z)},\,
p_z'\!+\!k_z\!+\!q_z,{\bf 0} \right) . \nn\\
\label{pprime}
\eea
Here $m_g \approx \mu/\sqrt 2 = gT \sqrt{N_c/3 + N_f/6}$ is the effective gluon mass
in a thermalized QGP at temperature $T$ with Debye screening mass $\mu$. 

The heavy quark energy loss per unit proper time $\tau$, to first order in opacity, can be
obtained by folding the heavy quark interaction rate $\Gamma(E)$ with the energy
$\omega + q_0$ and averaging over the initial color of the quarks 
\cite{Braaten:1991we,Le_Bellac} 
\bea
\frac{d E_{\mathrm{dyn}}}{d\tau}
= \frac{1}{D_R}  \int d\omega\, \omega \frac{d\Gamma(E)}{d\omega} 
\approx \frac{E}{D_R} \int dx\, x\frac{d\Gamma(E)}{dx} .
\label{dEdl}
\eea
The soft rescattering approximation $\omega + q_0 \approx \omega$ has been used, 
and $x$ is the longitudinal momentum fraction of the quark carried by the emitted gluon.
$D_R$ is defined as $[t_a,t_c][t_c,t_a] = C_2(G) C_R D_R$ with $C_2(G) =3$, $D_R=3$ 
and $[t_a,t_c]$ is a color commutator. The interaction rate is given by
\bea
\!\! \Gamma (E) &=& \frac{1}{2 E} 2\,{\rm Im}\,{\cal M}_\mathrm{tot} ,  \nn\\
\!\! &=& \frac{1}{2 E}\, (2\, {\rm Im}\,{\cal M}_{1,0} 
                    +2\, {\rm Im}\,{\cal M}_{1,1}
                    +2\, {\rm Im}\,{\cal M}_{1,2}),
\label{tot:int_rate}
\eea
which can be obtained by computing all the Feynman diagrams (see Appendix A-C)
that contributes at first order in opacity for the radiative energy loss.
${\cal M}_{1,0}$, ${\cal M}_{1,1}$ and ${\cal M}_{1,2}$ are the corresponding loop diagrams 
of the scattering amplitudes where, zero, one, two ends of the radiated gluon $k$ are attached 
to the exchanged gluon $q$ (see Figs. \ref{M10a:fig} and \ref{M12:fig} for illustration). 

In the high temperature plasma, the exchanged gluon receives
correction from medium partons. This many-body effects gets encoded in the
hard thermal loop (HTL) gluon propagators \cite{Kapusta}.
The effective 1-HTL gluon propagator has the form \cite{Kalash,Blaizot:2001nr}:
\bea
i D_{\mu\nu}(l) = \frac{P_{\mu \nu}(l)}{l^2 - \Pi_{T}(l)} + 
\frac{Q_{\mu \nu}(l)}{l^2 - \Pi_{L}(l)} ,
\eea
where the gluon-momentum $l = (l_0, \vec{\bf l})$. 
The usual ideal part for the longitudinal and transverse self energies 
\cite{Kalash,Blaizot:2001nr} are:
\bea
\Pi_T(l) &=& \mu^2 \left[ \frac{y^2}{2} 
+ \frac{y(1-y^2)}{4}\ln \left( \frac{y+1}{y-1} \right) \right], \nn\\
\Pi_L(l) &=& \mu^2 \left[ 1 - y^2 
- \frac{y(1-y^2)}{4}\ln \left( \frac{y+1}{y-1} \right) \right],
\eea
where $y \equiv l_0/|\vec{\bf l}|$. For typical values of 
$\eta/s \lesssim 3/(4\pi)$, required to explain flow data at RHIC and LHC,
small viscous corrections in $\Pi_{L,T}$ are obtained, which can be safely ignored.
The transverse $P^{\mu\nu}(l)$
and longitudinal $Q^{\mu\nu}(l)$ projector, in the Coulomb gauge, has the 
nonzero components $P^{ij}(l) =-\delta^{ij} +l^i l^j/\vec{\bf l}^2 = 1-y^2$.
The imaginary part of the exchanged gluon propagator is
\bea
D_{\mu\nu}^> (q) &=& (1+f(q))\; 2 \, {\rm Im} \left[ 
\frac{P_{\mu\nu}(q)}{q^2{-}\Pi_{T}(q)} + 
\frac{Q_{\mu\nu}(q)}{q^2{-}\Pi_{L}(q)}  \right] \nn\\
&&\times \theta\left(1-\frac{q_0^2}{\vec\bq^2}\right) .
\label{propagator}
\eea
The distribution function $f(q) = f_0(q) + \delta f_{\rm vis}(q)$ receives strong viscous 
correction $\delta f_{\rm vis}(q)$ due to Grad's 14-moment approximation of 
Eq. (\ref{fvis_Grad:eq}); the equilibrium gluon momentum distribution function 
$f_0(q) = [{\rm exp}(q_0/T) -1]^{-1}$.
For the radiated gluon, $\Pi_L(k) \approx 0$ and $\Pi_T(k) \approx m_g$.
Using the soft scattering limit ($\omega \gg |{\bf q}| \sim |{\bf k}| \sim gT$), 
and noting that $f(k) \ll 1$ for energetic partons \cite{Djordjevic:2007at}, 
the cut propagator for the imaginary part of the radiated gluon becomes
\bea
 \: D^>_{\mu\nu}(k) \approx - 2 \pi 
\frac{P_{\mu \nu}(k)}{2 \omega} \, \delta (k_0 - \omega) ,
\label{dmnk}
\eea
where $\omega \approx \sqrt{\vec\bk^2+m_g^2}$. The cut propagator for the heavy quark is
\bea
 \: D^>(p') \approx 2 \pi \, \frac{1}{2E'} \, \delta (p_0' - E') .
\label{dmnp}
\eea

With the help of these propagators one can calculate the matrix amplitude squared for
the diagrams (see Appendix A-C). The  phase space factor for the cut diagrams 
receives in-medium viscous corrections. 
On computing the diagrams, and in conjunction with Eqs. (\ref{dEdl}) and 
(\ref{tot:int_rate}), one can obtain the heavy quark radiative energy loss.
The contribution to the energy loss from the first set of diagrams, 
$2 \: {\rm Im} {\cal M}_{0,1}$, (see Eq. (\ref{AE10})) is given by
\bea
\frac{1}{E}\frac{dE}{d\tau}\Big|_{1,0} = \frac{12\alpha_s^2C_R T}{\pi}
\int dx \: d^2\bk \frac{\bk^2}{(\bk^2+\chi)^2} \Lambda (\tau) ,
\label{E10}
\eea
where $\chi = M^2x^2+m_g^2$, and the strong coupling constant $\alpha_s = g^2/(4\pi)$.
The medium informations are encoded within the quantity
\bea
\Lambda(\tau) &=& \! \int \! \frac{d^2\bq \:  dy}{(2\pi)^2} 
 \left( 1 + \frac{\Phi}{4sT^3}\frac{\bq^2(1-3y^2)}{1-y^2} \right) 
{\cal F}_{LT}(q) , \nn\\ 
&\equiv& \Lambda_0(\tau) + \delta\Lambda_{\rm vis}(\tau).
\label{Gradphi}
\eea
where $\Lambda_0(\tau)$ and $\delta\Lambda_{\rm vis}(\tau)$ stem from ideal and 
viscous correction due to Grad's 14-moment approximation (\ref{fvis_Grad:eq})
for $P=\epsilon/3$ equation of state. 
We have used the shorthand notation, 
${\cal F}_{LT} \equiv {\cal F}_L - {\cal F}_T$, for the difference 
of the polarization tensors
${\cal F}_Z = 2{\rm Im}\,\Pi_Z(y) 
[(\bq^2 + {\rm Re}\,\Pi_Z(y))^2 + ({\rm Im}\,\Pi_Z(y))^2]^{-1}$,
with $Z \equiv (L,T)$,
in terms of the dimensionless variable of Eq. (\ref{dimv}), viz.
$y=|\vec\bq|\cos\theta_q (\tau_0/\tau)[\vec\bq^2\cos^2\theta_q (\tau_0/\tau)^2+\bq^2]^{-1/2}$.
It is evident from the energy loss expression, that  
the nature of divergence gets modified from ideal to the viscous Bjorken case
due to an extra $\bq^2$ factor stemming from $\delta\Lambda_{\rm vis}$.

The diagram ${\cal M}_{1,2}$, where emission of a gluon occurs from the exchanged gluon,
has been computed in Appendix B. The corresponding radiative energy loss is given by
(see Eq. (\ref{BE12})):
\bea
\frac{1}{E}\frac{dE}{d\tau}\Big|_{1,2} = \frac{12\alpha_s^2C_R T}{\pi}
\int dx \: d^2\bk \frac{(\bk + \bq)^2}{[(\bk + \bq)^2 +\chi]^2} \Lambda(\tau) . \nn\\
\label{E12}
\eea

Finally, the diagrams for ${\cal M}_{1,1}$ can be computed as the product of the previous 
two diagrams ${\cal M}_{1,0}$ and ${\cal M}_{1,2}$. The resulting radiative energy loss gives
(Eq. (\ref{CE11}) in Appendix C):
\bea
\frac{1}{E}\frac{dE}{d\tau}\Big|_{1,1} \! = \frac{12\alpha_s^2C_R T}{\pi}  \!
\int \! dx \: d^2\bk \frac{-2 \bk \cdot (\bk + \bq)}{[(\bk + \bq)^2 + \chi][\bk^2 + \chi]} 
\Lambda(\tau). \nn\\
\label{E11}
\eea
The total energy loss is obtained by summing Eqs. (\ref{E10}), (\ref{E12}), (\ref{E11}) as
\bea
\frac{1}{E}\frac{dE}{d\tau}  &=& \frac{4\alpha_s C_R}{\pi\lambda_{\rm dyn}} 
\int dx \: d^2\bk  \nn\\
&& \times \left[ \frac{\bk}{\bk^2 +\chi}  
- \frac{\bk + \bq}{[(\bk + \bq)^2 +\chi]} \right]^2 \Lambda (\tau), \nn\\
&\equiv& \frac{4\alpha_s C_R}{\pi\lambda_{\rm dyn}} 
\int dx \: d^2\bk  \: {\cal P}_g(x,\bk,\bq) \: \Lambda(\tau) .
\label{totE}
\eea
In the above equation, a dynamical mean free path has been defined as 
$\lambda_{\rm dyn}^{-1} = C_2(G) \alpha_s T = 3\alpha_s T$.
In contrast to a time-independent QCD medium 
\cite{Djordjevic:2007at,Djordjevic:2008iz,Djordjevic:2009cr}
where $\lambda_{\rm dyn}$
is constant, in the present expanding medium $\lambda_{\rm dyn}$ and thereby the
density of scatterers has a time dependence via the temperature which modifies
the energy loss. 

Using the Chapman-Enskog results for the viscous evolution equation 
(\ref{CE:eq}) and the corresponding nonequilibrium distribution function 
(\ref{fvis_CE:eq}), the total radiative energy loss can be shown to have the 
same form as Eq. (\ref{totE}) in the Grad's approximation. 
However, in the CE method, the quantity $\Lambda$ 
of (\ref{Gradphi}) is replaced by
\bea
\!\! \Lambda(\tau) &=& \! \int \! \frac{d^2\bq \:  dy}{(2\pi)^2} 
 \left( 1 + \frac{5\Phi}{4sT^2}\frac{\bq (1-3y^2)}{1-y^2} \right) 
{\cal F}_{LT}(q) , 
\label{CEphi}
\eea
which involves the nonequilibrium form of the distribution function in the
CE method.

\section{Results and Discussions}

\begin{figure}[t]
\scalebox{0.4}{\includegraphics{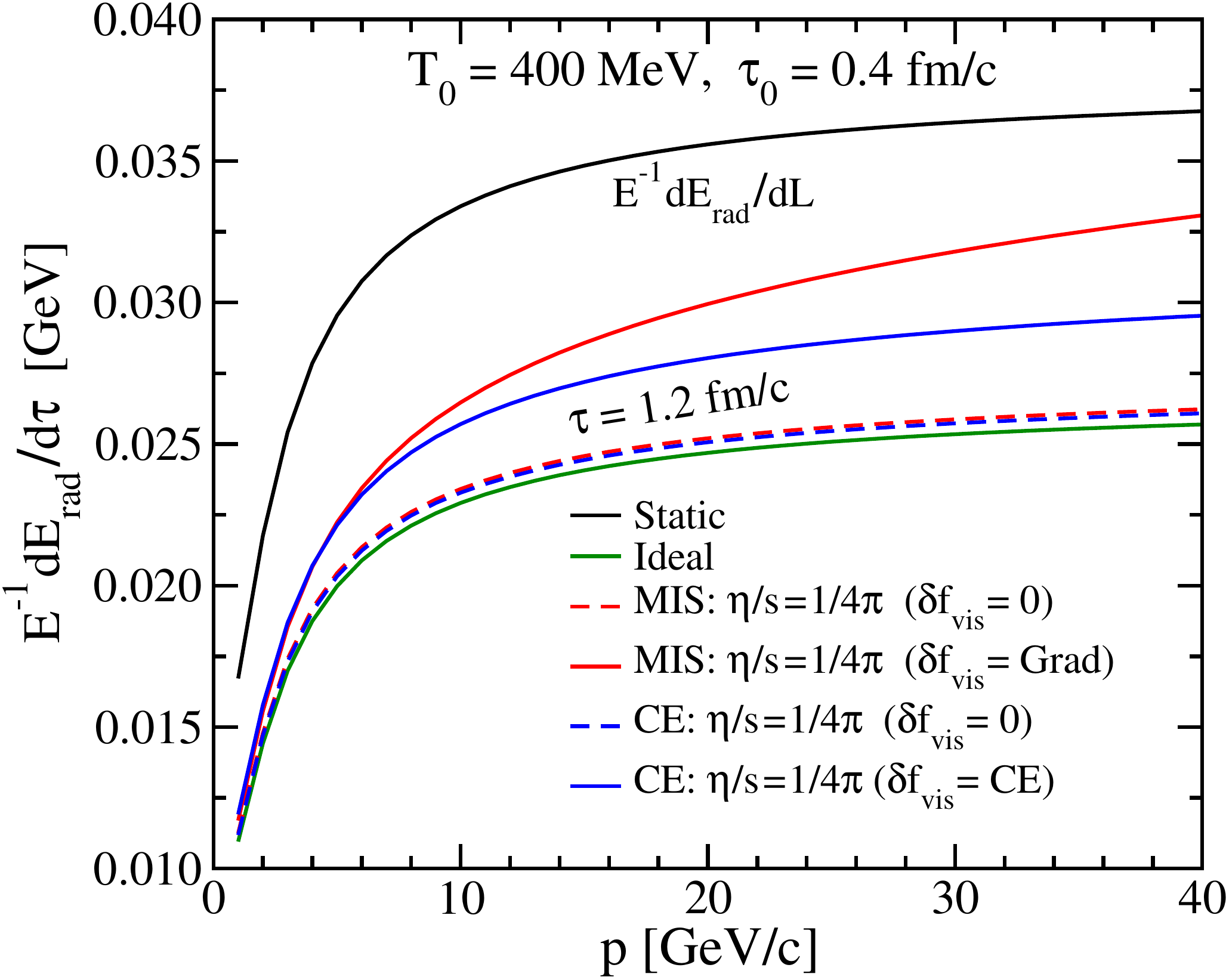}}
\caption{Fractional radiative energy loss as a function of momentum for charm quarks
in a boost-invariant expanding plasma at time $\tau = 1.2$ fm/c. The results 
are for ideal fluid (green solid line), in the dissipative hydrodynamics 
without viscous correction in M\"uller-Israel-Stewart (MIS) (red dashed line)
and Chapman-Enskog (CE) (blue dashed line) theories, and with further inclusion 
of viscous correction due to Grad in MIS (red solid line) and Chapman-Enskog 
(blue solid line) methods. The energy loss per unit length $dE/dL$ 
in a static plasma is also shown (black solid line). The results are for ideal gas equation 
of state ($P=\epsilon/3$) with initial temperature $T_0 = 400$ MeV, proper time 
$\tau_0 = 0.4$ fm/c, and constant shear viscosity to entropy density ratio $\eta/s = 1/4\pi$.}
\label{Emom:fig}
\end{figure}

In this section we estimate numerically the effects of expanding viscous medium
on the radiative energy loss in first order in opacity for a dynamically screened
QCD medium. We consider a plasma with an initial temperature $T_0 = 400$ MeV and
proper time $\tau_0 = 0.4$ fm/c that corresponds to (averaged) values obtained 
in Pb+Pb collision at LHC. Charm quark of mass $M = 1.2$ GeV is assumed to
traverse in the plasma that has an effective number
of degrees of freedom $N_f = 2.5$ with a constant strong coupling constant
$\alpha_s = g^2/4\pi = 0.3$ and Debye screening mass $\mu \sim gT$. 

Considering the case of an non-expanding static fluid at a constant temperature
$T_0 = 400$ MeV, the momentum dependence of fractional radiative energy loss 
$E^{-1} dE/dL$ of the charm quark is shown in Fig. \ref{Emom:fig} (black solid line)
\cite{Djordjevic:2007at}. In boost invariant 
longitudinal expansion of an ideal fluid, the temperature decreases with time
as $T = T_0 (\tau_0/\tau)^{1/3}$. 
This enforces a smaller fractional energy loss, $E^{-1} dE/d\tau$, of the charm quark 
as seen at a later time $\tau=1.2$ fm in Fig. \ref{Emom:fig} (green solid line).

With the inclusion of dissipation in the dynamical evolution, the temperature decreases 
at a slower rate and the entropy increases as compared to an inviscid fluid. 
In Fig. \ref{Emom:fig} we present the fractional radiative energy loss of charm quark
in an expanding viscous medium with $\eta/s = 1/4\pi$ at $\tau = 1.2$ fm/c in 
the MIS (red dashed lined) and
CE (blue dashed line) in absence of nonequilibrium part of the distribution function 
(i.e. $\delta f_{\rm vis} =0$). Dissipative effects is seen to cause
$\sim 5\%$ larger energy loss for charm quarks with momentum $p \geq 10$ GeV as compared
to that with ideal flow. Such an enhanced energy loss may be attributed to a
relatively higher instantaneous temperature of the viscous plasma. Although the
temperature in the CE approach falls slightly faster with time as compared to that 
in the MIS theory, the energy losses in these viscous evolution frameworks are 
found to be practically insensitive.

\begin{figure}[t]
\scalebox{0.4}{\includegraphics{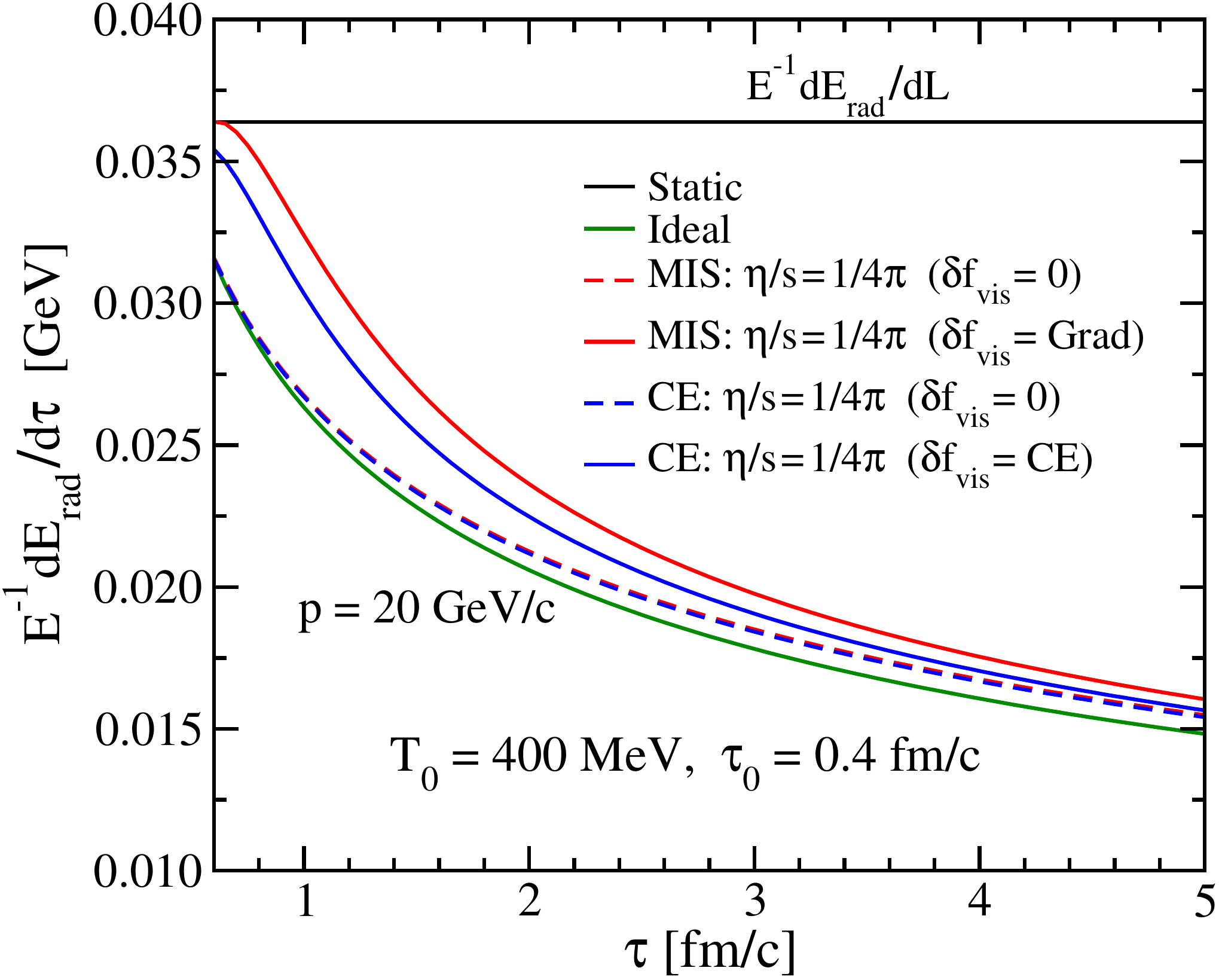}}
\caption{Time dependence of fractional radiative energy loss for charm quarks
of initial momentum $p = 20$ GeV/c. The initial conditions are
same as in Fig. \ref{Emom:fig}.}
\label{Etim:fig}
\end{figure}

Figure \ref{Emom:fig} also shows the fractional energy loss obtained by 
inclusion of viscous corrections in the single-particle distribution function using the 
Grad (red solid line) and Chapman-Enskog 
(blue solid line) methods at $\tau = 1.2$ fm/c. We find that nonequilibrium correction 
induces a significant increase in the energy loss;
albeit the magnitude of $E^{-1} dE/d\tau$ is still smaller than in a static fluid. 
The enhancement is particularly large for Grad's 14-moment approximation as compared to 
the Chapman-Enskog correction for heavy quark momentum $p \gtrsim 10$ GeV/c. This can be 
understood by comparing the (positive) contribution from viscous correction 
$\delta f_{\rm vis}$ to the energy loss in Grad and Chapman-Enskog approaches, 
namely, Eqs. (\ref{Gradphi}) and (\ref{CEphi}). 
An extra factor $\bq/5T$ in the integrand of $\delta\Lambda_{\rm vis}(\tau)$ in 
Grad's method gives a larger energy loss. Moreover, this energy loss is seen to rapidly 
increase with the momentum $p ~(= \sqrt{E^2-M^2})$ of the charm quark as the limit of integration 
$q_{\rm max} = \sqrt{4ET}$ \cite{Djordjevic:2003zk}. On the other hand, the energy loss
obtained in the Chapman-Enskog viscous correction shows a similar saturation pattern 
as that seen in an ideal fluid and in viscous medium with $\delta f_{\rm vis} = 0$.
At $p < 10$ GeV/c the energy loss has identical behavior for the
two viscous corrections used here. 
Large viscous corrections due to Grad's 14-moment approximation have been also found in 
the spectra and elliptic flow of hadrons at kinetic freezeout   
\cite{Romatschke:2007mq,Song:2007ux,Luzum:2008cw}, as well as in the
longitudinal Hanbury-Brown-Twiss-Radii radii of pions \cite{Bhalerao:2013pza}.

\begin{figure}[t]
\scalebox{0.4}{\includegraphics{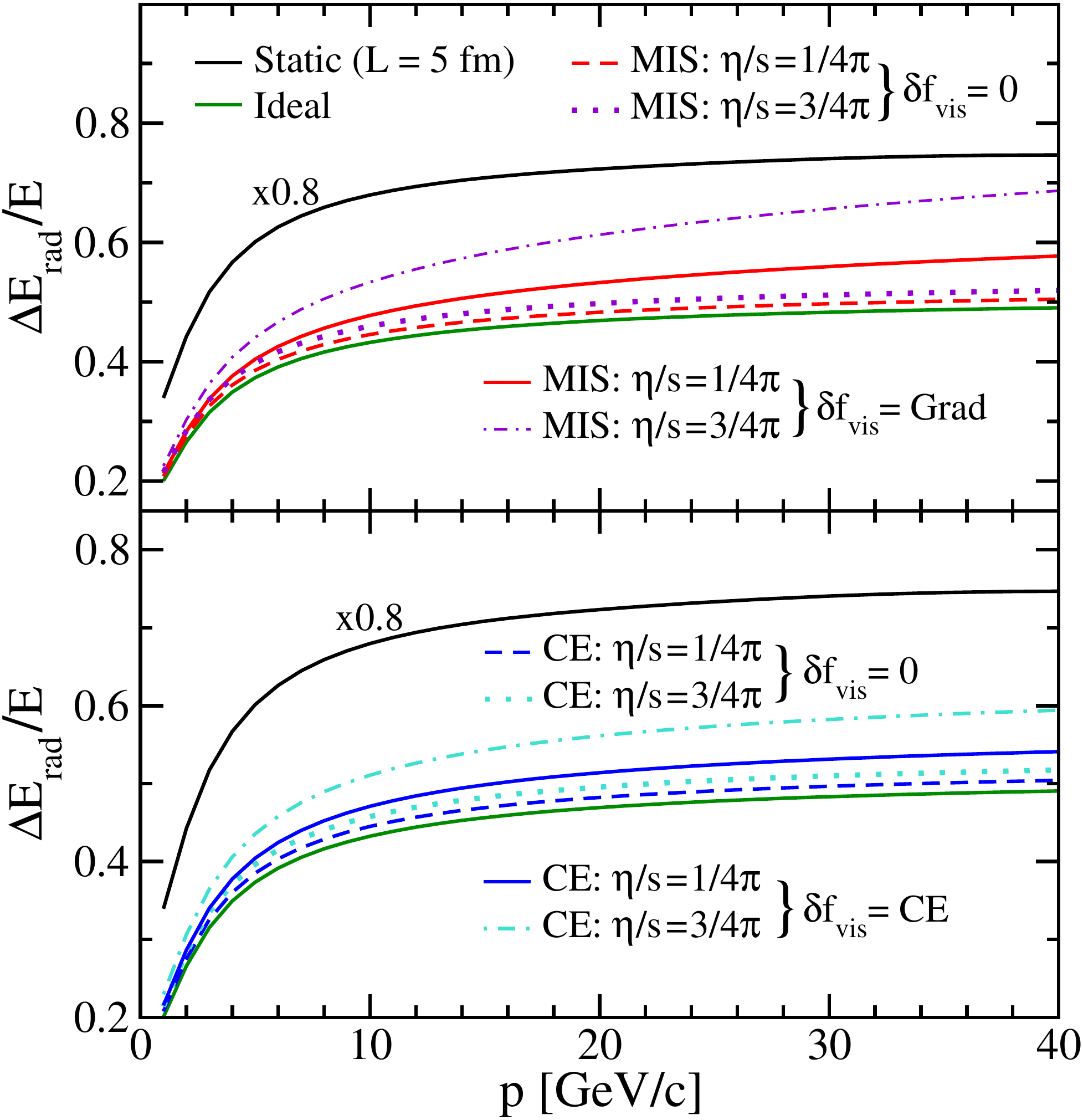}}
\caption{Time integrated fractional radiative energy loss as a function 
of momentum for charm quarks propagating in a boost-invariant expanding 
fluid over a total time of $\tau_f = 5$ fm/c with various values of $\eta/s$ in the
M\"uller-Israel-Stewart (MIS) (top panel) and Chapman-Enskog (CE) (bottom panel) 
frameworks. Also shown is the fractional energy loss in a static 
plasma integrated over a path length of $L =5$ fm.
The initial conditions are the same as in Fig. \ref{Emom:fig}.}
\label{Etot:fig}
\end{figure}

Figure \ref{Etim:fig} displays the proper time dependence of fractional radiative 
energy loss, $E^{-1} dE/d\tau$, for a charm quark of momentum $p = 20$ GeV/c.
As expected, the energy loss at all times in the expanding medium is smaller than in 
a static fluid. With increasing time, the decrease in the energy loss is essentially due to
falling temperature. As discussed above, we find viscous medium induces
a somewhat larger energy loss as compared to an ideal fluid at all times. 
Although at early times $\tau \lesssim 5$ fm/c,
the MIS dissipative hydrodynamics with viscous correction results in maximum energy loss,
at later times all the viscous fluids give nearly identical energy losses mainly due to
negligibly small shear pressure tensor in the dilute medium. Of course, for charm
quarks with momentum $p > 20$ GeV/c, the differences in $E^{-1} dE/d\tau$
will sustain at large times as evident from Fig. \ref{Emom:fig}.

\begin{figure}[t]
\scalebox{0.4}{\includegraphics{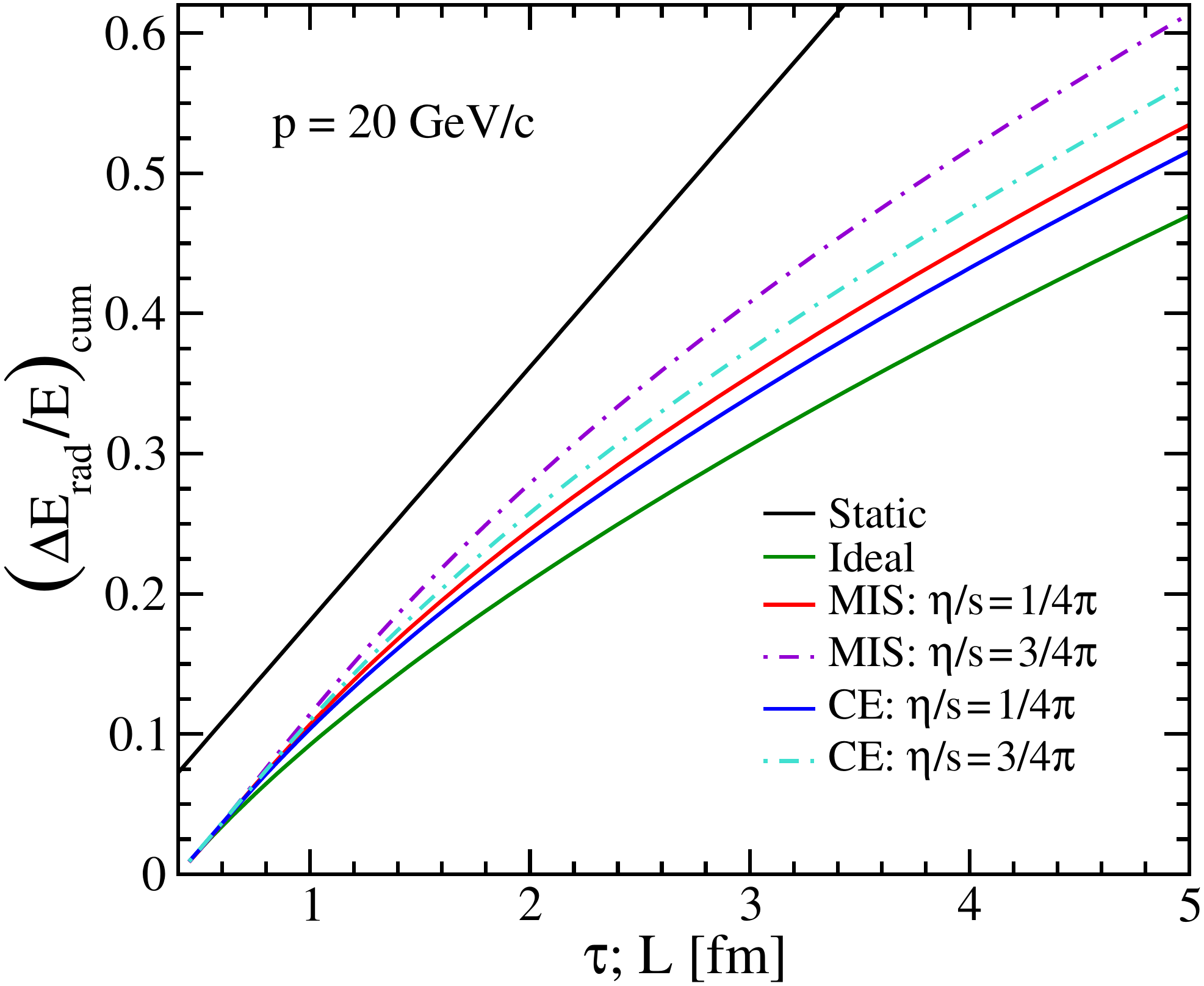}}
\caption{Time dependence of cumulative fractional radiative energy loss for charm quarks
of initial momentum $p = 20$ GeV/c with viscous corrections included in the distribution
function in models of dissipative hydrodynamics. The initial conditions are
same as in Fig. \ref{Etot:fig}.}
\label{Ecum:fig}
\end{figure}

We show in Fig. \ref{Etot:fig}, the charm quark momentum dependence of the (total) fractional 
energy loss $\Delta E/E$ at various values of $\eta/s$ in the MIS and CE theories. 
The total $\Delta E$ is obtained by summing the energy loss during the entire time 
traversed by the quark. In the present calculation we set this time as $\tau_f = 5$ fm/c, 
as the typical lifetime of the QGP phase at RHIC and LHC. On the other hand,
in a static fluid, $\Delta E$ refer to energy loss integrated over a path length of 
$L = 5$ fm. In an expanding medium the scattering rates decrease resulting in smaller 
$\Delta E/E$ as compared to the static case. However, large viscous corrections give 
positive contribution to the energy loss that grows with $\eta/s$, especially when 
nonequilibrium part of distribution function in the Grad's approximation is considered. 
In fact, the energy loss results obtained in various dissipative hydrodynamic medium 
fall between those in the static fluid and ideal hydrodynamics.

 \begin{figure*}[t]
 \scalebox{0.36}{\includegraphics{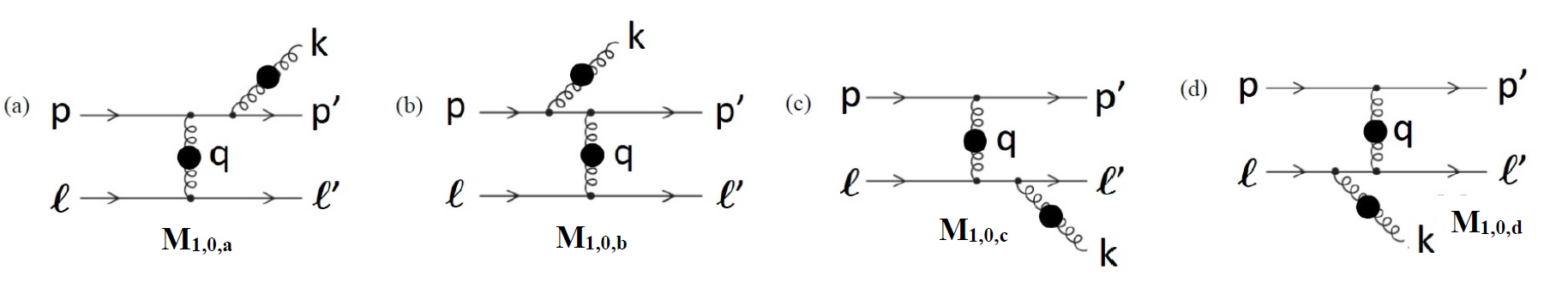}}
 \caption{Feynman diagrams $M_{1,0}$ (a)-(d) 
contributing to heavy quark radiative energy loss to first order in opacity.}
 \label{M10_all:fig}
 \end{figure*}
 

It is important to note that in a static medium, the total energy loss 
can be expressed as $\Delta E \sim {\cal S}(E) \int dL/(3\alpha_s T)$,
where ${\cal S}(E)$ is a function of the energy of charm quark. 
Hence the total energy loss increases linearly with the path length $L$ traversed by the quark
which is QCD analogue of QED Bethe-Heitler limit \cite{Baier:1996sk}.
On the other hand, the corresponding energy loss for a time-dependent viscous medium 
may be written as $\Delta E \sim \int d\tau \: {\cal C}(E,\tau,\eta)/ (3\alpha_s T(\tau))$,
where ${\cal C}(E,\tau,\eta)$ is a complicated function encompassing medium effects.
Figure \ref{Ecum:fig} illustrates the time dependence of cumulative energy loss in the
ideal and dissipative hydrodynamics. The corresponding energy loss for a static medium is 
shown as a function of effective thickness of the medium. We find that expanding  
medium shows a non-linear behavior in the cumulative energy loss which has been also 
observed in coherent gluon radiation from static medium \cite{Djordjevic:2003zk}.

\section{Summary}

In this work we have presented a theoretical formulation of the radiative energy loss 
of heavy quark traversing in a viscous medium that undergoes boost-invariant longitudinal 
expansion. The calculation was performed in first order in opacity for a dynamically 
screened QCD plasma at finite temperature. We have derived the radiative energy loss 
by including two forms of viscous correction in the nonequilibrium phase-space distribution,
namely the Grad's 14-moment approximation and the Chapman-Enskog-like iterative solution. 
The evolution of the medium was treated within relativistic second-order viscous hydrodynamics
based on  M\"uller-Israel-Stewart (MIS) framework, that uses Grad's approximation for distribution 
function, and Chapman-Enskog (CE) method.
Viscous contributions from dynamics only, in absence of viscous corrections in the single-particle
phase-space distribution, resulted in the enhancements of the fractional energy loss energy by 
about $\sim 5\%$ depending on the shear viscosity to entropy density ratio of 
$\eta/s = 0.08 - 0.24$ used. This energy loss was found to be similar in the MIS and CE 
dissipative hydrodynamic models. 
At the early stages of evolution, we found that inclusion of Grad's approximation of viscous correction 
in the distribution function resulted in appreciably large increase of fractional energy loss 
that increased monotonically with momentum $p$ of the charm quark.
On the other hand, in the Chapman-Enskog viscous correction, the enhancement was found to be 
comparatively smaller, and the energy loss was seen to saturate for $p \gtrsim 10$ GeV/c. 
At later proper times, the energy losses
in all the scenarios were found comparable due to small temperature and nearly vanishing
shear stress tensor. The time integrated fractional energy loss in the Grad's approximation was 
found higher than in the Chapman-Enskog method.
The heavy quark radiative energy loss results presented in this work is crucial 
for the interpretation of D-meson nuclear modification factor.

\appendix

 
 \begin{figure}[b]
 \scalebox{0.2}{\includegraphics{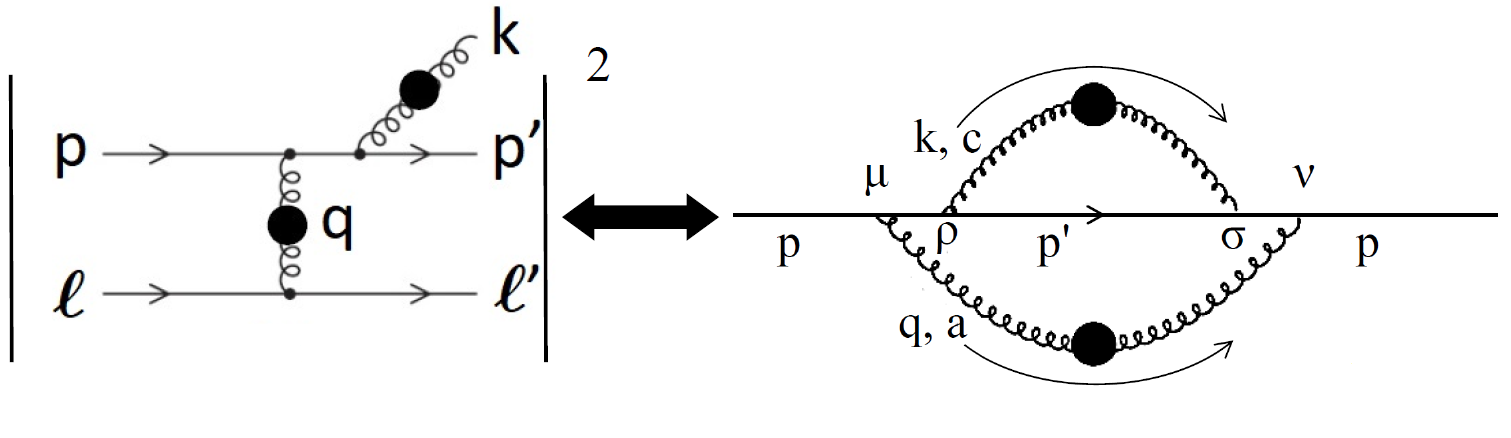}}
\caption{Left: Scattering amplitude of $M_{1,0,a}$ diagram where a heavy quark of momentum $p$ 
suffers collisional interaction with medium partons via screened gluon of momentum $q$,
resulting in emission of a gluon of momentum $k$ from outgoing quark. 
The blob represents medium modified gluon propagator. 
Right: HTL loop diagram of first order in opacity corresponding to ${\cal M}_{1,0,a}$. 
Heavy quark scatters from medium partons via cut gluon propagator of 
momentum $q$ (with $q_0 \leq |\vec\bq|)$ resulting in emission of a cut gluon propagator with momentum 
$k$ (with $\omega > |\vec\bk|$). The imaginary part of the diagram corresponds to the squared amplitude 
of the left diagram and integrated over phase-space.}
\label{M10a:fig}
\end{figure}


\section{Computation of diagrams ${\cal M}_{1,0}$ and associated radiative energy loss in MIS theory}

We present detailed calculation of the first set of diagrams corresponding to ${\cal M}_{1,0}$. 
In general, we denote all the loop diagrams as ${\cal M}_{1,i,j}$, where $i$ refers to the number of the
exchanged gluon $q$ that are attached to the radiated gluon $k$, and $j=a,b,\ldots$ denotes the
particular diagram in that class, computed in first order in opacity denoted by $1$.
The Feynman diagrams for the first set, namely
$M_{1,0,a}$, $M_{1,0,b}$, $M_{1,0,c}$, and 
$M_{1,0,d}$ are shown in Fig. \ref{M10_all:fig}.  
These scattering diagrams are associated with two-cut HTL loop diagrams.
We first compute the cut diagram ${\cal M}_{1,0,a}^> = 2{\rm Im} \ {\cal M}_{1,0,a}$
(see Fig. \ref{M10a:fig})
\bea
{\cal M}_{1,0,a}^> &=&g^4t_a t_c t_c t_a \int 
\frac{d^4p^{'}}{(2\pi)^4}\frac{d^4q}{(2\pi)^4} \frac{d^4 k}{(2 \pi)^4}\nn\\
&&(2p{-}q)^\mu    D^>_{\mu \nu}(q) (2p{-}q)^\nu \left[D(p'+k)\right]^2 \nn\\
&&\times (2p'{+}k)^\rho D^>_{\rho\sigma}(k) (2p'{+}k)^\sigma
   D^>(p')\nn\\
&&\times (2\pi)^4\delta^4\left(p - p' - k - q \right).
\label{int_rate:1}
\eea      
The above equation consists of two parts: the medium interaction and the phase space factor. 
The interaction history is encoded in the exchanged and radiated gluon propagators 
$D_{\mu \nu}(q)$ and $D_{\rho\sigma}(k)$, respectively; $D(p')$, 
$D(p'+k)$ are the fermionic propagators. 
To proceed further we write the vector contraction as
\bea
&&(2p' + k)^\rho P_{\rho\sigma}(k) (2p' + k)^\sigma  \nn\\
&&\approx  2p'^\rho P_{\rho\sigma}(k)2p'^\sigma 
\approx - 4\left(\vec{\bf p'}^2 - \frac{(\vec{\bf p'}
\cdot \vec{\bf k})^2}{|\vec{\bf k}|^2}\right),
 \label{pre_fac:M1}
 \eea
where we have used $k^\rho P_{\rho\sigma}(k)=0$.
By choosing the coordinate axis
$\vec\bq= |\vec\bq|(\sin\theta_q\cos\phi_q,\sin\theta_q \sin\phi_q,\cos\theta_q)$,
$\vec\bk= |\vec\bk|(\sin\theta_k\cos\phi_k,\sin\theta_k \sin\phi_k,\cos\theta_k)$ and  
$\vec{\bf p}'$ along $z$ direction, 
one can evaluate the terms within the braces of Eq.(\ref{pre_fac:M1}) as 
\bea
\vec{\bf p'}^2 - \frac{(\vec{\bf p'} \cdot \vec{\bf k})^2}{|\vec{\bf k}|^2} &\approx&
\frac{p_z'^{2}\bk^2 }{\bk^2+p_z'^2x^2} \approx \frac{\bk^2}{x^2} ,
\label{vec1_eq}
\eea
where $x =: k_z/p'_z$. Similarly, for the vector contraction with the exchanged gluon term 
one can write 
\bea
p^\mu \,  {\rm Im} \: P_{\mu\nu}p^\nu \approx 
- \frac{E^2\bq^2}{\bq^2 + q_z^2} 
\approx- p^\mu \,  \, {\rm Im} \: Q_{\mu\nu}p^\nu .
 \eea
Other approximations which we use are $q_z{\,\sim\,}|{\bf q}|$, ${\bf |k|}\ll k_z$, 
$q_z{\,\ll\,}k_z$. The longitudinal component of the emitted and radiated gluons obeys the following approximations,
 $k_z{+}q_z \approx k_z$, $p'_z{+}k_z{+}q_z \approx p'_z{+}k_z \approx p'_z$ and 
$p'_z+q_z\approx p'_z$. For the energy delta function we thus obtain,
   \bea
   \delta(E-E'-\omega-q_0) \approx \delta(q_z-q_0).
  \eea
 While writing the above equation it has been assumed that 
$M^2/2p_z'^2\ll 1,~ ({\bf k}^2{+}m_g^2)/2 k_z\ll 1,~ ((\bk+\bq)^2+M^2)/2 p_z'\ll 1$. 
Similarly, for the propagator one can write,
  \bea
  (p'+k)^2-M^2 &=&2 \left( p'_z+\frac{(\bk+\bq)^2+M^2}{2p'_z}\right) \nn\\
 && \times \left(k_z+ \frac{\bk^2 + m_g^2}{2 k_z}\right)\nn\\
 && -2\left[k_z p'_z + \bk \cdot (-\bk-\bq)\right], \nn \\
  &\approx& \frac{\bk^2 + M^2x^2 + m_g^2}{x}.
 \label{prop_eq}
  \eea
By using Eqs. (\ref{pre_fac:M1})-(\ref{prop_eq}), 
along with Eqs. (\ref{dmnk})-(\ref{dmnp}), the Eq. (\ref{int_rate:1}) reduces to
  \bea
{\cal M}_{1,0,a}^> \!&=&\! 16g^4t_a t_c t_c t_a \! \int \! 
\frac{dp_0'}{2\pi}\frac{d^4q}{(2\pi)^4} \frac{d^4 k}{(2 \pi)^4}
  \frac{\bk^2}{(\bk^2+M^2x^2+m_g^2)^2}  \nn\\
&& \times (1+ f_q)\, \frac{E^2\bq^2}{\bq^2+q_z(\tau)^2} \Bigg\{
2{\rm Im} \left( \frac{1}{q^2{-}\Pi_{L}(q)}  \right) \nn\\
&& -  2{\rm Im} \left(\frac{1}{q^2{-}\Pi_{T}(q)}\right)\Bigg\}
2\pi \frac{\delta (p_0'-E')}{2 E'} \,
2 \pi \, \frac{\delta (k_0 - \omega)}{2 \omega} \nn\\
&& \times 2\pi\delta(p_0{-}p_0'{-}k_0{-}q_0) \:
\theta\!\left(1-\frac{q_0^2}{\vec{\bf q}^2} \right).
\label{int_rate:M1a}
\eea
In presence of viscous correction due to Grad (\ref{fvis_Grad:eq}) and in-medium modifications 
\cite{Moore:2004tg}, the bosonic distribution function becomes
\bea
f(q) &=& f_0(q)+\frac{3\Phi}{4sT^3} \Big[ \frac{\bq^2 + q_z^2(\tau_0/\tau)^2}{3} \nn\\
&& - q_z^2\frac{\tau_0^2}{\tau^2} \, f_0(q)(1+f_0(q)) \Big],
\label{dist_func:1}
\eea
where $q_z = |\vec\bq| \cos\theta_q$.
In the high temperature plasma and small $q_0$, the equilibrium part of the distribution function 
can be approximated as
$f_0(q)(1+f_0(q))\simeq f_0(q)\simeq 1/(1+ q_0/T - 1)\simeq T/q_0=T/q_z$.
Since $q^2=q_0^2-q_z^2-\bq^2$, and using the delta function, we can write
$q_0\sim q_z$, $q^2\approx-\bq^2$.  
To proceed, we have introduced a dimensionless variable
$y=q_0/q \sim q_z(\tau)/q$, which can be also written as
\bea
y&=&\frac{|\vec\bq| \cos\theta_q (\tau_0/\tau)}{\sqrt{\vec\bq^2\cos^2\theta_q (\tau_0/\tau)^2+\bq^2}}.
\label{dimv}
\eea
Limits on $y$ are decided by $\cos\theta_q$ viz.  $y \in [y_{\rm min},y_{\rm max}]$. 
On performing the $p_0$, $k_0$ and $q_0$ integrations, we finally
get,
\bea
{\cal M}_{1,0,a}^> &=& 8g^4t_at_ct_ct_a ET\int  \frac{d^3 k}{(2 \pi)^32\omega}
  \frac{\bk^2}{(\kperp^2+M^2x^2+m_g^2)^2} \nn\\
&&\times \int \frac{\bq d\bq \ dy \, d\phi}{(2\pi)^2}
\left(1+\frac{3\Phi}{4sT^3}\frac{\bq^2(1-3y^2)}{3(1-y^2)} \right) \nn\\
&& \times \Bigg\{\frac{2{\rm Im}\,\Pi_L(y)}
 {(\bq^2 + {\rm Re}\,\Pi_L(y))^2 + ({\rm Im}\,\Pi_L(y))^2}\nn\\
&& - \frac{ 2{\rm Im}\,\Pi_T(y)}
{(\bq^2 + {\rm Re}\,\Pi_T(y))^2 + ({\rm Im}\,\Pi_T(y))^2} \Bigg\}.
\label{M011}
\eea
It can be shown that contribution from the other three
diagrams, ${\cal M}^>_{1,0,b}$, ${\cal M}^>_{1,0,c}$, ${\cal M}^>_{1,0,d}$, has the same result but for 
the color factor. On summing all the four diagrams and using Eqs. (\ref{dEdl}) 
and (\ref{tot:int_rate}), the heavy quark radiative energy loss 
with the Grad's viscous correction for this set:
\bea
\frac{1}{E}\frac{dE}{d\tau}\Big|_{1,0} &=& \frac{2 g^4 T[t_a,t_c][t_c,t_a]}{(2\pi)^5 \: D_R}
\int_0^1 \! dx \int_0^{k_{max}} \! \bk d\bk  \int_0^{2\pi} \!  d\phi_k \nn\\
&& \times \int_0^{q_{max}} \!\! \bq d\bq   \int_0^{2\pi} \!  d\phi_q
\int_{y_{\rm min}}^{y_{\rm max}} \! dy  \nn\\
&& \times \frac{\bk^2}{(\bk^2+M^2x^2+m_g^2)^2} \nn\\
&& \times \left( 1 + \frac{\Phi}{4sT^3}\frac{\bq^2(1-3y^2)}{1-y^2} \right) 
{\cal F}_{L,T}. 
\label{AE10}
\eea
With the help of the commutator relation, $[t_a,t_c][t_c,t_a] = 3 C_R D_R$,
and defining the strong coupling constant $\alpha_s = g^2/(4\pi)$,
the coefficient in front of the integral can be written as 
$3\alpha_s^2 C_R T/\pi^3$.
We use the notation ${\cal F}_{LT} \equiv {\cal F}_L - {\cal F}_T$ for 
the difference of the polarization tensors
${\cal F}_Z = 2{\rm Im}\,\Pi_Z(y) 
[(\bq^2 + {\rm Re}\,\Pi_Z(y))^2 + ({\rm Im}\,\Pi_Z(y))^2]^{-1}$,
with $Z \equiv (L,T)$. The upper limits of integration are set to 
$q_{\rm max} = \sqrt{4ET}$ and $k_{\rm max}=2E\sqrt{x(1-x)}$
\cite{Djordjevic:2003zk}.

\section{Computation of diagram ${\cal M}_{1,2}$ and corresponding radiative energy loss}

 \begin{figure}[t]
 \scalebox{0.21}{\includegraphics{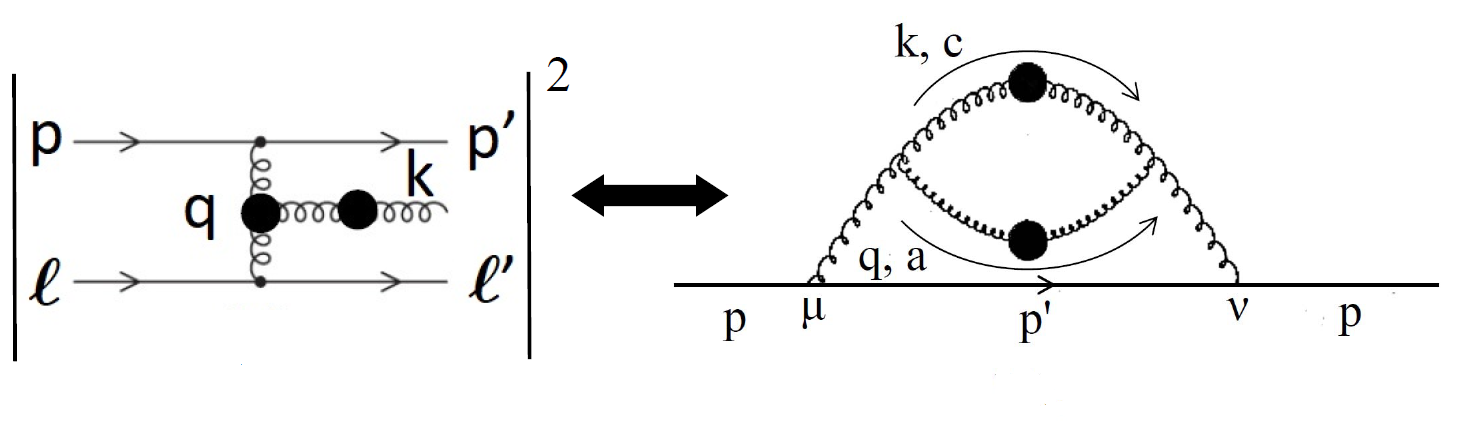}}
\caption{Feynman diagram $M_{1,2}$ for heavy quark radiative energy loss to 
first order in opacity (left), and the corresponding loop diagram ${\cal M}_{1,2}$ (right).
The notations are same as in Fig. \ref{M10a:fig}, except that the emission 
of gluon of momentum $k$ occurs from the virtual/exchanged gluon of momentum $q$.}
 \label{M12:fig}
 \end{figure}

We present detailed calculations for diagram corresponding to ${\cal M}_{1,2}$ where
both ends of the exchanged gluon $q$ are attached to the radiated gluon 
(see Fig. \ref{M12:fig}). The contribution of this diagram is given below,
${\cal M}_{1,2}^> = 2{\rm Im} \, {\cal M}_{1,2}$:
\bea
{\cal M}^>_{1,2} &=& \frac{g^4}{2E}\,f^{bac} t_b f^{dac} t_d 
 \int \frac{d^4p'}{(2\pi)^4} \frac{d^4q}{(2\pi)^4}\,\frac{d^4k}{(2\pi)^4}\, \nn\\
 && \times (2\pi)^4\delta^4(p{-}p'{-}k{-}q)D^>(p')\, H, 
\label{M_2q}
\eea
where we have defined
 \bea 
  H&=& (2p{-}k')^\mu\,(2p{-}k')^\nu\ D_{\mu\rho}(k')\,D^>_{\lambda\alpha}(k)\,
       D^>_{\tau\beta}(q)\,D^*_{\sigma\nu}(k') \nn \\
 && \times \Bigl(g^{\rho\tau}(k'{+}q)^\lambda + g^{\lambda\tau}(k{-}q)^\rho
     - g^{\lambda\rho}(k'{+}k)^\tau\Bigr) \nn \\
 && \times \Bigl(g^{\sigma\beta}(k'{+}q)^\alpha + g^{\alpha\beta}(k{-}q)^\sigma
          - g^{\alpha\sigma}(k'{+}k)^\beta\Bigr).
\eea
We follow similar algebra for vector contraction as used in Eqs. (A2)-(A6),
The fermionic propagator can be expressed as 
\bea
(k{+}q)^2-m_g^2 &=&m_g^2-\bq^2 + 2\left(k_z+\frac{\bk^2+m_g^2}{2k_z}\right) \nn\\
&& \times \left(q_z-\frac{\bk^2 + M^2x^2 +m_g^2}{2k_z}\right) \nn\\
&& -2k_z q_z-2\bk\bq - m_g^2 , \nn\\
&\approx& -[(\bk{+}\bq)^2 + M^2 x^2 +m_g^2] .
\eea
Further, using $if^{bac}t_b = [t_a,t_c]$ and the viscous correction due to Grad 
(see Eq. (A8)), one can compute the diagram of Eq. (\ref{M_2q}).
The corresponding radiative energy loss in Grad's 14-moment approximation:
\bea
\frac{1}{E}\frac{dE}{d\tau}\Big|_{1,2} &=& \frac{2 g^4 T [t_a,t_c][t_c,t_a]}{(2\pi)^5 \: D_R}
\int_0^1 \! dx \int_0^{k_{max}} \! \bk d\bk  \int_0^{2\pi} \!  d\phi_k \nn\\
&& \times \int_0^{q_{max}} \!\! \bq d\bq   \int_0^{2\pi} \!  d\phi_q
\int_{y_{\rm min}}^{y_{\rm max}} \! dy  \nn\\
&& \times \frac{(\bk+\bq)^2}{[(\bk+\bq)^2+M^2x^2+m_g^2]^2} \nn\\
&& \times \left(1 + \frac{\Phi}{4sT^3}\frac{\bq^2(1-3y^2)}{1-y^2} \right) 
{\cal F}_{L,T}. 
\label{BE12}
\eea


\section{Computation of diagrams ${\cal M}_{1,1}$ and corresponding radiative energy loss}

We present calculations of the diagrams ${\cal M}_{1,1}$ where
one of the ends of the exchanged gluon $q$ is attached to the radiated gluon.
This can be evaluated as the product of the previous two diagrams. For the
first diagram ${\cal M}^>_{1,1,a}$ one can express 
\bea
{\cal M}^>_{1,1,a} &\approx& \frac{g^4}{2E} f^{cba} t_b t_c t_a 
 \int \frac{d^4p'}{(2\pi)^4}\frac{d^4q}{(2\pi)^4} \frac{d^4k}{(2\pi)^4}\nn\\ 
&& \times \frac{1}{(p'{+}k)^2-M^2-i \epsilon}  \nn \\
&& \times (2\pi)^4\delta^4(p{-}p'{-}k{-}q) D^>(p') \, G ,
\label{M_1q}
 \eea
where we denote 
\bea
G &\approx&  
\bigl[(2p-k')^\mu (2p'+k)^\nu(2p-q)^\sigma \nn\\ 
&& \times D_{\mu\rho}(k') D^>_{\nu\lambda}(k)\,D^>_ {\sigma\tau}(q)\bigr]\nn\\
&&\times \Bigl(g^{\rho\tau}(k'{+}q)^\lambda + g^{\lambda\tau}(k{-}q)^\rho
   - g^{\lambda\rho}(k'{+}k)^\tau\Bigr) , \nn \\
&\equiv& G_1 + G_2 - G_3.
\eea
Here,
\bea
G_1&=&\bigl[(2p{-}k')_\mu D^{\mu\rho}(k')D^>_{\rho\sigma}(q)\,
(2p{-}q)^{\sigma} \bigr] \nn\\
&&\times \bigl[(k'{+}q)^\lambda D^>_{\lambda\nu}(k)\,(2p'{+}k)^\nu \bigr] , \\
G_2&=&\bigl[(2p{-}k')^\mu D_{\mu\rho}(k')\,(k{-}q)^\rho\bigr] \nn\\
&&\times \bigl[(2p'{+}k)^\nu D^>_{\nu\lambda}(k) D^>_{ \, \lambda\sigma}(q)\,
            (2p{-}q)_{\sigma}\bigr] , \\
G_3 &\approx&  
  \bigl[(2p-k')_\mu D^{\mu\rho}(k') D^>_{\rho\nu}(k)\,(2p'+k)^\nu\bigr]\nn\\
&& \times \bigl[(k{+}k')^\tau D^>_{\tau\sigma}(q)(2p{-}q)^\sigma\bigr] .
\eea
We consider only $G_3$ as it gives a dominant contribution in the approximations 
involving the kinematics noted in Appendix A. With the help of the above equations and 
viscous correction Eq. (A8), one can compute
the energy loss for the  diagram ${\cal M}^>_{1,1,a}$. The energy loss for the other diagrams
in this set, ${\cal M}^>_{1,1,b}$, ${\cal M}^>_{1,1,c}$, ${\cal M}^>_{1,1,d}$, can be 
calculated accordingly. 
On summing all the four diagrams we get the total energy loss for this set as 
\bea
\frac{1}{E}\frac{dE}{d\tau}\Big|_{1,1} &=& \frac{2 g^4 T [t_a,t_c][t_c,t_a]}{(2\pi)^5D_R}
\int_0^1 \! dx \int_0^{k_{max}} \! \bk d\bk  \int_0^{2\pi} \!  d\phi_k \nn\\
&& \times \int_0^{q_{max}} \!\! \bq d\bq   \int_0^{2\pi} \!  d\phi_q
\int_{y_{\rm min}}^{y_{\rm max}} \! dy  \nn\\
&& \times \frac{-2\bk \cdot (\bk+\bq)}{[(\bk+\bq)^2+M^2x^2+m_g^2][\bk^2+M^2x^2+m_g^2]} \nn\\
&& \times \left( 1 + \frac{\Phi}{4sT^3}\frac{\bq^2(1-3y^2)}{1-y^2} \right) 
{\cal F}_{L,T}. 
\label{CE11}
\eea


\end{document}